# ABIOGENESIS: A POSSIBLE QUANTUM INTERPRETATION OF THE TELEPOIETIC CONJECTURE

Vittorio Cocchi*, Rossana Morandi**

*ATI, Roma
**Università degli Studi di Firenze, Dipartimento di Energetica

## Abstract

*In the research on the origin of life, topics that can be considered reasonably shared by the generality of researchers are initially identified. It is then shown that the application of these principles to the results obtained with the IdLE-IdLA mathematical model for the simulation of aggregative processes, leads to the conclusion that the primordial formation of self-replicating structures is difficult to reconcile with deterministic aggregative dynamics in the classical sense. Regardless of the extent to which the process itself is governed by chance or by aggregative codes written in the laws of chemistry, no conventional causality is likely. The model itself suggests only one possible way out, consistent with thermodynamics: the existence of information sets rushing into the system in a different way from the perceived time stream. The possibilities offered by quantum mechanics and its most recent interpretations are consequently investigated to try to interpret, at the level of particle physics, the suggestion of IdLE-IdLA model. The attempt leads to a mutual accreditation of macroscopic telepoiesis and a kind of quantum retrocausality. The result is a vision of the natural world in which the coexistence of causal and retrocausal dynamics is presented as a possible interpretative key of the whole complex of vital manifestations.*

## 1. Hypotheses about the birth of life

It is well known that the first appearance of life on Earth from non-living matter, the so-called *abiogenesis*, is not yet fully explained. The problem, in essence, is to justify how, starting from the simple organic compounds likely to be present in the atmosphere or in the waters of the early Earth, agglomerations with the minimum requirements to be considered living have been able to form and evolve: that is, how protocells have acquired the ability to use materials in the environment to maintain and renew their structure (*metabolic function*) and to reproduce themselves (*self-replication*). Many are the hypotheses proposed but much less are the experimental findings: in fact, no organic picture is currently shared by the scientific community. Science has been able to reconstruct in a precise and reliable way the first moments of the Universe (which no one has attended) but is not yet able to propose a convincing explanation on the appearance of life which so closely affects our world and ourselves. Similarly, Science is not yet able to explain in a complete and convincing way the biological evolution whose dynamics are only partly justified by that Darwinian paradigm that seems now lacking to an increasing number of researchers, at least in its ultraorthodox form: from Niles Eldredge to Stephen Jay Gould, from Steven Peter Rose to Richard Lewontin and many others.

A historically very followed group of patterns is based on the hypothesis that the formation of protocells equipped with a primitive form of metabolism is even antecedent to the onset of genetic information. It has been observed, in fact, that aqueous solutions of two or more polymers (protein, glucidic or lipid) segregate tiny droplets rich in colloids, called *coacervates*, with interesting structural and functional properties. Delimited by a pseudo-membrane, they are able to attract some macromolecules (polypeptides and polynucleotides) and to reject others.[1] Recent experiments have shown that some of these coacervates (...but with proper biochemical support!) are able to carry out something similar to a photosynthesis and can undergo a sort of natural selection as a result of their primitive metabolic abilities. Even outside the aqueous environment, pseudo-cellular structures have been obtained: starting from amino acids, suitably arranged on hot lava rocks (to simulate possible physical conditions on the primordial Earth), the formation of polymer chains similar to proteins has been observed. It has also been experimentally demonstrated that a subsequent watering (representative of primordial rains) tends to organize this organic material into microspheres (a kind of protocells) coated by a semipermeable membrane and equipped with a rudimentary metabolism, as they are capable of assimilating similar protocells. However, they are completely devoid of genetic endowment and are therefore unable to replicate and evolve.[2]

In order to overcome this not small problem, one of the most followed lines of research currently bases the birth of life on the initial spontaneous formation of primordial RNA molecules.[3] It has been shown that molecules of this type can have an autonomous enzymatic activity and, therefore, can be able to self-replicate using one of their internal subunits as a mould. The idea is that, starting from a primitive RNA world, increasingly complex self-replicating molecules would evolve, due to competition for the environment and the consequent natural selection, to generate the current world where DNA and proteins perform the essential genetic and metabolic functions to life as we know it. This hypothesis, however, leaves many doubts of both chemical and probabilistic nature. In fact, it has been demonstrated in the laboratory that mixtures of nucleotides can actually give rise to RNA molecules capable of self-replicating, mutating and competing with each other ...but only in the presence of protein enzymes![4] No researcher has yet been able to explain how significant quantities of these molecules may have formed spontaneously in the atmosphere or in the primordial oceans, since their synthesis is still complicated, uncertain and completely not reproducible in the laboratory. Nor is it clear how these macromolecules may have been sufficiently stable over time to have had the opportunity to evolve themselves. Much is still studied on the subject but the underlying difficulties remain: the hypothesized primordial RNA molecules remain in any case too complex structures for their random formation in a sufficiently large number of stable specimens to be considered probabilistically reasonable.

Without going into further detail, we can say that research on the origin of life continues on many fronts, but we can also say that we are far from being able to draw convincing scenarios. The recurring obstacle with which all the hypotheses are usually collided is the probabilistic criticality of the proposed models. All theories, in fact, in one way or another always end up assuming that chance, in spite of its intrinsic blindness, is capable of having decisive ordering effects: but as soon as one applies to the calculations, realizes how the possibility that structures of biological interest can be formed by chance without the guidance of an assembling code is statistically remote. Fred Hoyle said that the probability that random chemical processes have resulted in the creation of life is the same as a whirlwind hitting a landfill site and creating a plane.

---

[1] The first to follow this idea was the Russian biochemist Alexander Oparin (1894-1980).

[2] The forerunner of this line of research was the American biochemist Sidney Walter Fox (1912-1998), whose first experiments date back to the late 1970s.

[3] The first to make such a hypothesis was the American biochemist Thomas Robert Cech (1947-...), who was awarded the Nobel Prize in Chemistry in 1987.

[4] Studies of this type were conducted by the German chemist Manfred Eigen (1927-2019), winner of the Nobel Prize in Chemistry in 1967, and by the British chemist Leslie Orgen (1927-2007).

Regardless of the hypotheses proposed in the context of the researches just outlined (and in many other collateral lines of research) we believe that some common topics can be identified, probably shared by anyone approaching abiogenesis:

- the birth of life on Earth was evidently allowed by a natural tendency to the aggregation of elementary entities in more organized structures
- after its first appearance, the evolution of life has followed (and follows) rough paths, tortuous and not without errors and failures, but all of them with the same tension towards a growing complexity of organisms
- the prebiotic material must have been sufficiently abundant to make possible a wide variety of combinations before obtaining appropriate structures capable of self-replicating and initiating an evolutionary process
- since the molecules capable of guiding the aggregative processes in accordance with modern biochemistry (enzymes and nucleic acids) are themselves highly evolved, the initial aggregation of inorganic matter to the goal of self-replication can only have happened spontaneously without the help of external ordering agents.

It is in this context that the IdEP-IdLA mathematical model, designed specifically to study the chemistry of a generic aggregative process (and whose conceptual structure will be briefly recalled in the following paragraph), can be useful. The model, in fact, is able to highlight both the thermodynamic conditions which the reactions that govern the formation of complex structures must generally underlie and the probability that such structures manifest properties of biological interest such as self-replication.

## 2. The IdEP-IdLA model

The acronym IdEP-IdLA[5] refers to a mathematical model [4,5,6,7] based on the use of ideal elementary particles (IdEPs) of linear shape: each of them is characterized by a mass, a moment of inertia and specific aptitudes for coupling described, for a particle system, by a matrix of reciprocal binding energies. As a result of chemical reactions, differently governed by chance or necessity, the model is able to simulate the formation of macromolecular chains (IdLAs) in both closed and open thermodynamic systems. IdEPs can be regarded as monomers (similar, for example, to amino acids or nucleotides), and IdLAs, in turn, can be considered as the result of a polymerization process (something similar to proteins or ribonucleic acids).

The core of the model is the Markovian expression of the molar entropy of a gaseous mixture:

---

[5] IdEP is the acronym of *Ideal Elementary Particle*; IdLA is the acronym of *Ideal Linear Aggregate*.

$$S = \frac{5}{2}RlnT + R\left[ln\frac{V}{n_A} - lnv^* + 1 + h\right] \qquad (1)$$

where $R$ is the gas constant, $T$ is the absolute temperature, $V$ is the volume occupied by the mixture, $N_A$ is the Avogadro number, $v^*$ is the volume of the allocation cell[6] calculated as a weighted average of the allocation cell volumes attributable to the individual chemical species present in the mixture, $h$ is the entropy of the mixture descriptor, interpreted as Markov source.

Formula (1) is obtained by inserting elements derived from information theory in the well-known formula of entropy, proposed by statistical mechanics for linear molecules [16]. This particular expression of entropy allows to set the evolutionary study of an IdEPs system by conditioning the process itself as a consequence of the degree of constraint with which the individual couplings are expected to take place.

In particular, when (1) is used to describe the entropy of a mixture that is the product of an aggregative process (in progress or at the equilibrium, in both closed and open systems), the entropy of the product descriptor $h_{IdLA}$ takes the following expression

$$h_{IdLA} = (1 - \eta)h_0 \qquad \text{(with } 0 \leq \eta \leq 1)$$

where $h_0$ is the entropy of IdLA descriptor when couplings occur randomly and $\eta$ is the coding factor that determines the degree of conditioning to which the evolution of the system is subject when the aggregation is influenced by elements guiding the process more or less strongly. The null value of $\eta$, implying $h_{IdLA} = h_0$, then describes a mode of aggregation which, as characterized by a random sequence of elementary particles, refers to the behaviour of an information source *without memory*. A value of $\eta$ between 0 and 1 implies, instead, that the aggregation process is affected by factors (internal or external to the reagent system) which, influencing the formation of the aggregates in various ways, cause the process itself to refer to the behaviour of a *with memory* information source. Evidently, the unit value of $\eta$, implying $h_{IdLA} = 0$, identifies an entirely forced process.

Thus, (1) makes it particularly clear how an aggregative process can be interpreted as the implementation of a program by an information set that operates on the process itself. The model, in its theoretical development, allows this information to be placed both *inside* or *outside* the reagents system. In the first case the information lies in the matrix that defines the binding energies between pairs of particles, as normally happens in the elementary chemical reactions: we speak in this case of *autopoietic* process. In the second case, the information is placed in ordering agents which, although they do not provide material, guide the synthesis process, as happens in the formation of large polymers of biological interest:[7] in this case we speak of *heteropoietic* process. Fig.1 summarizes how, within the IdEP-IdLA model, the coding factor affects autopoietic aggregation and heteropoietic aggregation and which natural processes can be implicitly referred to.

The model, which can be operated under a large number of conditions, produces results perfectly in line with what is expected from the thermodynamics of both equilibrium and non-equilibrium states for aggregative reactions.[8] In particular it clearly shows that a heteropoietic process always implies that, at the equilibrium (at the end of the reaction in closed systems or at the achievement of a stationary condition in open systems), the free energy of Gibbs for the system (reagents and products together) is higher than what the system would have if it had been left free to evolve autopoietically.

---

[6] The *volume of the allocation cell* is the portion of space within which the position of a generic particle (elementary, i.e. IdLE, or produced by aggregation, i.e. IdLA) must be defined to be consistent with the quantum assumptions of statistical mechanics: for further details, see [4].

[7] The biological polymers implicitly referred to are proteins (in which the serial assembly of amino acids is promoted by enzymes as biocatalysts) and ribonucleic acids (as holders of the assembly code).

| PROCESS | Coding factor | Outcome of the aggregation process | Exemples in nature |
|---|---|---|---|
| **AUTOPOIETIC** *(due to the differentiation of bonding energies)* | $\eta = 0$ | Completely random | --- |
| | $0 < \eta < 1$ | Partially random | Partially probabilistic reactions |
| | $\eta = 1$ | Totally deterministic | Almost all general chemistry reactions |
| **HETEROPOIETIC** *(due to the action of external ordering agents)* | $0 < \eta < 1$ | Partially random | Partially probabilistic reactions |
| | $\eta = 1$ | Totally deterministic | The aggregative processes of natural polymers |

*Fig.1 – IdEP-IdLA model domain*

The graph in Fig.2, where apexes *a* and *h* refer, respectively, to autopoietic aggregation and heteropoietic aggregation, makes this conclusion evident. In particular, the diagram shows that the reaction enthalpy $\Delta H_R$ (negative, as the reaction is assumed to be exothermic) is always such that $\left|\Delta H_R^h\right| < \left|\Delta H_R^a\right|$ and the reaction entropy $\Delta S_R$ (also negative as the process normally generates order) is always such that $\left|\Delta S_R^h\right| > \left|\Delta S_R^a\right|$. Therefore, at the equilibrium the system undergoing a heteropoietic aggregation has a higher free energy than the same system undergoing an autopoietic aggregation: that is, in the first case a surplus of order, represented in the graph by the *not compensated entropy* $\sigma$, is observed. The entropic balance is satisfied only if it is extended to the dissipative processes related to the activity of the ordering agents themselves.[9]

This is a less trivial result than it may seem. The IdLE-IdLA model, in fact, is not limited to pointing out that the more information is fed into a system, the more its entropy decreases but rather shows how, in the competition between an implicit assembly code and one imposed from the outside, the prevalence of the latter necessarily requires a compensation mechanism for the entropy deficit. It also allows us to continuously and at will weigh the contribution of chance and necessity through the parameter $\eta$. For clarification, an example is proposed to which explicit reference will be made in paragraph 5.

[8] The model, in fact,
- at the equilibrium, gives reason for the incompleteness of some reactions, indicates the causes and responds correctly to the sensitivity analysis
- in stationary non-equilibrium conditions, presents an entropy production, due to internal aggregative processes, that always balances the variation of entropy due to thermal and material flows. It also matches the principle of minimum entropy production
- far from equilibrium, convincingly describes dissipative structures.
[9] In open systems the model also demonstrates the relevance of the catalytic action of the ordering agents to compensate σ: an increase in the reaction speed determines, in stationary conditions, an increase in the production of entropy.

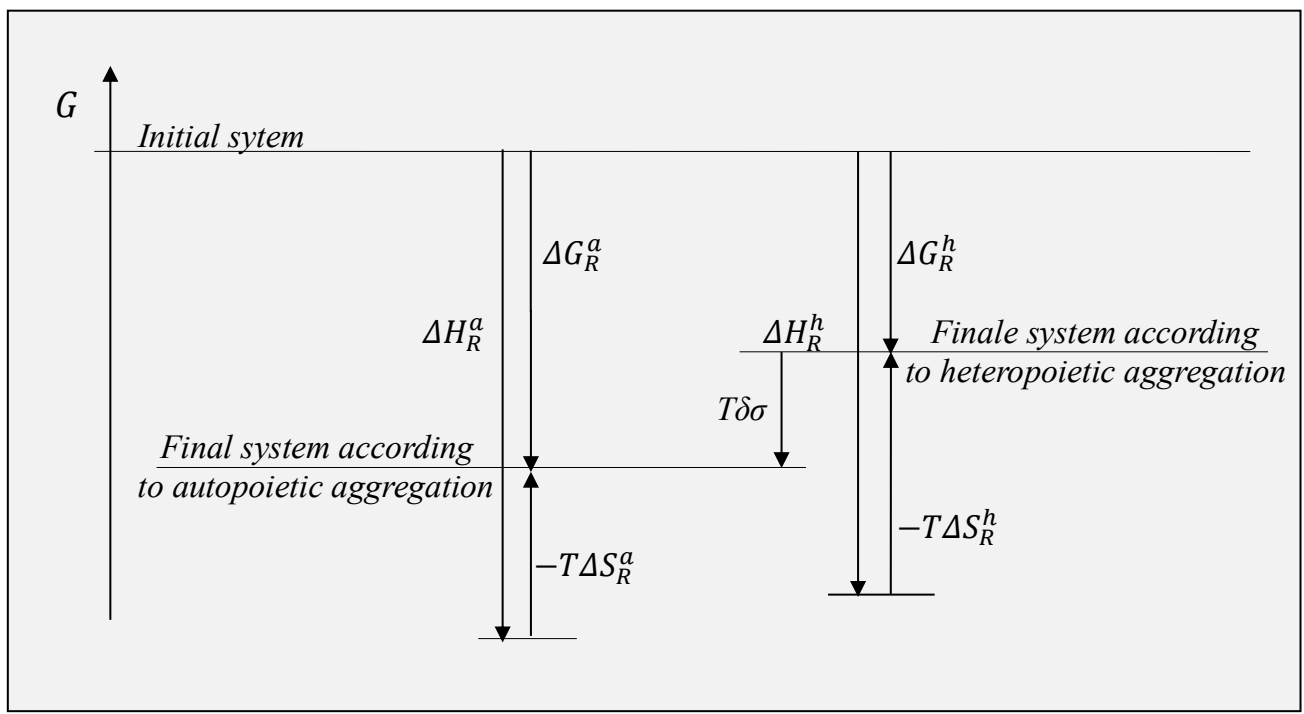

*Fig. 2 – Qualitative comparison between the energy evolution of a system that aggregates in an autopoietic way and the energy evolution of the same system subject to the action of external ordering agents*

Let us suppose that an initial system consists of a mixture in which four different types of elementary particles (IdEP1, IdEP2, IdEP3, IdEP4) are present in the same proportions and that conditions are favourable to the occurrence of an aggregative process. Let us also suppose that a given IdLA, formed for example by a predetermined sequence of 100 IdEPs, is of special importance from an abiogenetic point of view: with reference to what is discussed in paragraph 1, let us assume that it shows self-replicating properties. We call this particular IdLA *target aggregate*.[10]

If couplings are random[11] ($\eta = 0$), the entropy $h_{IdLA}$ of the product descriptor (whose sequencer is represented by a no-memory information source) and the probability $P_0$ of obtaining a target aggregate are respectively

$h_{IdLA} = h_0 = \sum_{i=1}^{4} p_i ln\frac{1}{p_i} = 1,386$ where $p_i = 0,25$ with $i$ ranging from 1 to 4

$P_0 = 0,25^{100} = 6,223 \ 10^{-61}$

---

[10] The following arguments are substantially correct but simplified: for a formally correct theoretical treatment see the bibliography [4,5,6,7].

[11] In chemical terms, this refers to the case where there is no differentiation in the binding energies between different IdEP pairs.

If instead, each time, the sequencer regularly proposes the right IdEP ($\eta = 1$), then the entropy $h_{IdLA}$ of the product descriptor (whose sequencer is represented, this time, by a source of information with a properly deep memory) and the probability of obtaining the target aggregate are respectively

$$h_{IdLA} = h_1 = 0$$
$$P_1 = 1$$

Obviously, for values of $\eta$ between 0 and 1, the probability that the sequence is formed, each time, by the right IdEP increases, passing gradually from a condition of total randomness to a condition of absolute necessity.

The graph of Fig.3 shows the trends of the coding factor $\eta$, the entropy $S$ of the final system and the probability $P$ to obtain the target aggregate, depending on the probability $p_{yes}$ that the correct IdEP is sequenced each time (having assumed, for simplicity, that the three incorrect IdEPs have the same probability to occur). This shows that particularly high coding factors are not needed to obtain a significant number of target aggregates: $\eta = 0,7$ gives one target aggregate per 20.000, $\eta = 0,8$ one per 280 and $\eta = 0,9$ one per 11. Such aggregation frequencies can definitely be considered of biological interest.

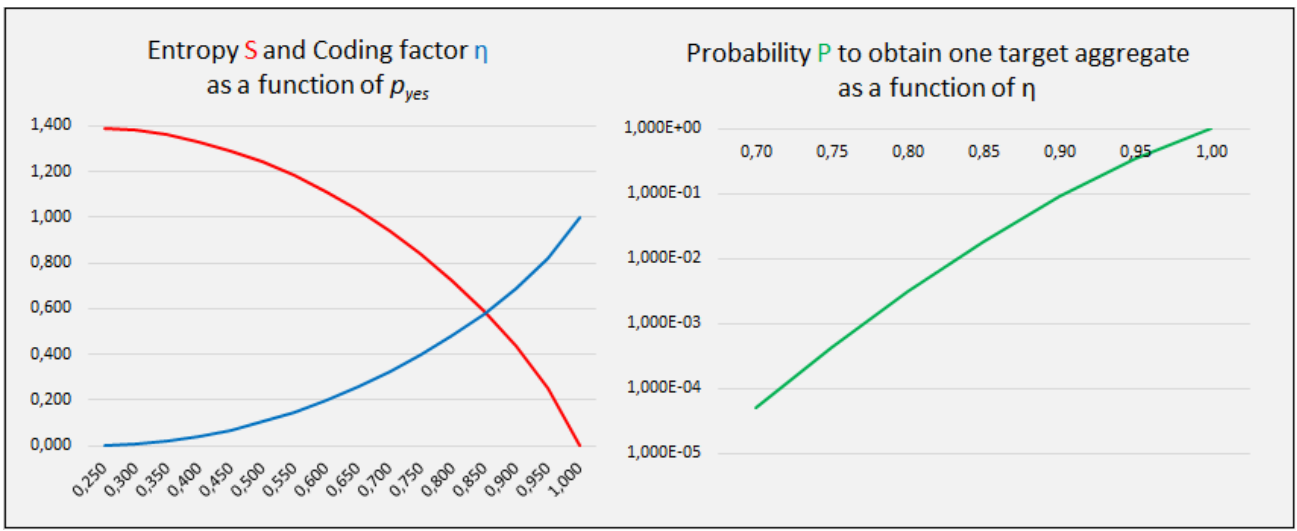

*Fig.3 – Target aggregate formation: entropy, coding factor and probability depending on the probability $p_{yes}$ that the correct IdLE occurs each time.*

But the model also provides another important indication, which we will refer to later: in open systems far from equilibrium, the ordering agents must not only ensure that the IdEP chains are sequenced so that the aggregates reach the expected levels of organisation, but must also exert appropriate catalytic action. Only in this way, in fact, the maximum production of entropy is reached and therefore the stability [7]. This is also a not trivial result, as it allows, among other things, to analyse the conditions under which the dissipative structures of Prigogine [24] in the chemical field must be subjected.

### 3. Abiogenesis interpreted with the IdEP-IdLA model

We now propose to interpret the results obtained with the IdEP-IdLA model in an abiogenetic context, in the light of the four points proposed at the end of paragraph 1.

First, it is clear that the primordial aggregative process that gave rise to life cannot have been heteropoietic: the ordering agents (enzymes and nucleic acids) are themselves too complex and specialized structures to admit their existence in a prebiotic scenario. The problem would only shift.

But even the autopoietic coded aggregation does not appear to be a viable hypothesis. The strong causal component theoretically underlying this hypothesis would imply, in fact, that the material world, starting from its deeper characterization (like the fundamental constants) contains in itself rules oriented to the construction of an adequate number of increasingly complex structures until the appearance of life. The ability to favour the formation of *fertile* compounds at the expense of others should therefore constitute a sort of program, written in depth in the matter. But if autopoiesis is, to some extent, acceptable for the formation of the first organic compounds, it is less and less a reasonable mechanism, as we proceed in the direction of biocomplexity. Otherwise, it should be admitted that in the conformation of atoms and in the elementary interactions between them and within them, from the very beginning of the history of the universe, there is a code of exceptional power by virtue of which, step by step, life must blossom as a fact of necessity. A code not only powerful, not only pervasive but also exceptionally targeted because among all the possible paths, the matter is aggregated following a particularly virtuous logic.[12] We must not forget that, among all the possible amino acid sequences, those that form useful proteins are incredibly few. The same can be said for the serial assembly of nucleotides: it is practically impossible to obtain an operable nucleic acid if not assembling the base material according to exact modalities. It is therefore hardly plausible that the enormous variety and even greater complexity of everything around us, including ourselves, are the result of extraordinarily successful combinations, managed by a single code that has guided nature for billions of years. In conclusion, autopoiesis is a mechanism too critical to be credible: even if this is an aggregative process that certainly exists on a small scale, it cannot alone justify the biocomplexity that we observe.

Thus, if it is objectively difficult to accept that the aggregative dynamics underlying abiogenesis may have been based on chance, or may have been aided by unlikely pre-existing external ordering agents, or have been promoted by an autopoietic code carved *ab aeterno* in the matter, what suggestions does the model IdEP-IdLA provide?

If we are inclined to overcome the limitations of the traditional dialectic tension between chance and necessity, the model reveals a possible way out: the existence of dynamics that provide flows of information into the process in ways not attributable to the classic concept of causality in accordance with the arrow of time usually perceived.

---

[12] The wonder caused by the extreme criticality of such a scenario is summed up in the expression *fine tuning* and leads to acknowledge how special our world is, starting from the specific values characterizing the constants that regulate the fundamental physical interactions. One of the philosophical consequences is the so-called *anthropic principle* in all its different forms. On the other hand, the alternative to the acknowledgment of the extraordinary uniqueness of our physical world is that infinite worlds exist and that we live precisely in that one that has coincidentally assumed the configuration, among all the possible ones, that allowed it to generate ourselves, thinking beings, reflecting on its own nature. It is the *multiverse theory*, which postulates the existence of other universes, in parallel dimensions. This is an idea first proposed at the end of the 50s of the last Century, which has some basis in quantum mechanics and also results as a possible consequence of unifying theories such as string theory. However, since it is not a falsifiable hypothesis, it is today regarded with suspicion, if not with sufficiency, by many researchers.

In fact, if the heteropoietic mode is not admissible, in a traditionally causal perspective, for the unlikely presence of ordering agents, this seems to become immediately acceptable if an information flow that streams into the system according to an inverted time arrow or even to an atemporal mode is invoked to compensate the entropy deficit $\sigma$ of Fig.2. We then call *telepoiesis* this particular mode of aggregation, due to its manifestly finalistic nature and *not compensated neghentropy* the entropy deficit $\sigma$ which, in the context of classical thermodynamics, remains unbalanced.[13]

But if the attempt to interpret abiogenesis in the light of the IdEP-IdLA model, leads us to suppose that there may be chemical-physical processes whose macroscopic results seem justifiable only by invoking non-conventional flows of information, we wonder: can all this be explained in terms of quantum theory?

In reality, some interpretations of quantum mechanics admit the existence of backword in time and/or atemporal processes at the quantum substratum level, thus giving space to information flows that might rush into an evolving system in ways other than those consistent with a one-way flow of time. We then aim to understand whether the indications provided by our model can be justified in terms of such peculiar interpretations of quantum mechanics and what could the term *telepoiesis* really mean in this context.

The following paragraph is intended to give a brief overview of the interpretations of quantum mechanics, functional to our purpose.

4. Retrocausality: a pervasive suggestion

It is still controversial whether the fundamental laws of nature, expressed by quantum mechanics and combined in the standard model of elementary particle physics, are time-in-variant. That is, whether the asymmetry of time, so evident and pervasive in our daily life, is to be considered directly derived from a temporal asymmetry rooted in the profound reality of things, or is not rather due to our subjective perception of events, ordered according to that direction of time implying an increase in the measured entropy.[14]

The subject is widely debated [13,25], also because the epistemological implications are extremely significant: to the extent that the quantum-mechanical (relativistic) model could be considered reversible (at least in its deterministic component, that is, before the collapse of the wave function), then extending the temporal symmetry of the fundamental laws to causal dependencies, at least in the domain of the Shroedinger equation, would seem quite natural. And if so, it would be possible to assume that, at the fundamental level, causal influences (and therefore flows of information) back in time no less than forward in time can take place. It is evident how such a hypothesis can affect the idea that we might form about the type of objective reality staying beyond the predictive and explanatory success of modern quantum mechanics; so much so that, since the middle of the last century, many researchers have developed interpretative models of quantum mechanics that leave room for the existence of causal influences that operate backwards in time.

---

[13] The concept of negative entropy was first introduced by Erwin Schrodinger in 1943 in the popular book "What is life?". The synthetic term *neghentropy* was later coined by French physicist Léon Brillouin to indicate the accumulation of negative entropy in living beings. As an alternative to this term, the synonym *syntropy* was later proposed in 1942 by the Italian mathematician Luigi Fantappié in his "Unitary theory of the physical and biological world" and also adopted in 1974 by the Hungarian physiologist Albert Szent-Gyorgyi.

[14] The *deterministic part* of non-relativistic quantum mechanics, expressed by the Schroedinger equation in which a time derivative of the first order of $\Psi$ is present, is not symmetrical with respect to time: symmetry is recovered only if, in addition to the transformation t → -t (adopted in the classical context), the complex conjugation of the wave function $\Psi$ (necessary to obtain the right values of energy and momentum) is also introduced. This additional condition, although operationally justified, is considered by many physicists not to legitimate quantum reversibility in a non-relativistic quantum-mechanical context. On the other hand, in the same Shroedinger equation rewritten taking into account the special relativity (at least in the case of a single particle), we find a time derivative of the second order and consequently, in this case, temporal symmetry would be claimed. Thus, even if the *non-deterministic* part of quantum mechanics, represented by the collapse of the wave function $\Psi$, describes such a causal discontinuity that temporal symmetry cannot be admitted in any way, many believe that *before collapse* the state of the system can be characterized by substantial reversibility.

The first proposal of *retrocausality* in quantum mechanics was suggested by Wheeler and Feynman [26,27] who, between 1945 and 1949, proposed their own interpretation of the wave equation derived by Dirac to describe the quantum-relativistic motion of electrons: equation that allows solutions with positive energy but also with negative energy. Acknowledging that this equation implies a substantial invariance of the described phenomenon with respect to a temporal inversion, the two physicists constructed a mathematical model to study the physical behaviour of an accelerated charged particle, interpreting it as a process of interaction between a source and an absorber, rather than as an independent elementary process. They thus identified two possible solutions of the electromagnetic field equations: a first solution, called *delayed*, which, with respect to the established arrow of time, comes from the past and a second solution, called *advanced*, which, instead, would come from the future.[15] The gratifying symmetry that the model of *advanced and delayed waves* represents, though conceived in the middle of the last century, has been for a long time, and still is, the most important reference for the retrocausal models subsequently developed in the context of quantum mechanics. Among these, the so-called *transactional model*, due to the US physicist John G. Cramer [9,10,11,12], is worth mentioning for the particular favour with which it was accepted and further developed.

According to this theory, real physical events are identified by what Cramer calls *handshakes* between quantum states that evolve both back and forth in time. According to Cramer, when a vibrating electron (or an excited atom) emits a quantum of energy (a photon), the source produces a kind of *offer wave*. Similar to Wheeler-Feynman's interpretation, this field propagates in space both back and forth in time. If this field encounters an absorber, a *confirmation* wave is generated which also propagates both back and forth in time, thus acting on the emitter at the very moment of emission. The delayed field produced by the absorber and the anticipated field produced by the emitter are cancelled respectively with the delayed field produced by the emitter and the advanced field produced by the absorber, for all times prior to the emission of the photon and after its absorption. The only recordable radiative field is between the emitter and the absorber. It is important to note that the theory admits that more emitters and more absorbers can be present at the same time: but only transactions complying with quantum boundary conditions are allowed. Any observer witnessing this process would perceive only the completed transaction, interpretable, in fact, as a handshake (a bond) between emitter and absorber. Thus, in terms of the usual quantum mechanical formalism, transactional theory interprets the wave function as an offer wave and the collapsed wave as the completion of the transaction.

---

[15] Wheeler and Feynman imagined that an electromagnetic signal, associated with the acceleration of a charged particle, initially travels towards the absorber, perturbing its individual particles, and that the absorber, in turn, generates an electromagnetic response field. According to their view, this field would be composed of the sum of half of the delayed solutions (forward in time) and half of the advanced solutions (back in time) of Maxwell's equations. In particular, the sum of the advanced effects of all the absorber particles would produce an advanced field arriving at the charged particle simultaneously with the emission. This advanced field would exert a finite force on the source that has exactly the amplitude and direction required to account for the energy transferred from the source to the absorber: the observed effect would thus be the Dirac radiation damping. When this advanced field is combined with the equivalent half-delayed and half-advanced field of the source, the observed total disturbance is the entire delayed field known empirically to be emitted by the accelerated charged particle. Because of the advanced field of the absorber, the radiative damping field would thus be present at the source exactly at the time of the initial acceleration.

But in very recent studies (the oldest of which [18] dates back to 2012), some researchers have subjected Cramer's theory to a careful epistemic analysis: according to them, the transaction, with its offer and confirmation waves in space-time, implies the existence of hidden variables that propose a deterministic interpretation of phenomenological reality. The world that would result would be a block world, substantially inconsistent with a dynamically open vision of reality: that is, a world in which transactions are basically predefined without leaving room for possibility. Therefore, retrocausality, understood as a flow of information from the future to the past, constitutes a narrative valid from the formal point of view but presenting problems of consistency such that it cannot aspire to describe any ontological reality. On the other hand, it is more consistent and promising to assume that Cramer's transactions are taking place in a pre-temporal dimension, without hidden variables. In this way it can be admitted that an information set, living in a timeless domain, breaks into a physical system in a non-deterministic way, establishing its macroscopic effects in spacetime. It is a scenario that fits very well to interpret the telepoiesis suggested by our model and to which the authors are inclined to adhere. We will return to this in the next paragraph.

Meanwhile, it is worth concluding this brief review by illustrating how the idea of retrocausality has played a pervasive role even in fields of research that go well beyond basic physics.

In fact, the suggestions induced by quantum retrocausality have led many researchers to export this concept from the microscopic world of quantum to the world of macroscopic phenomena. One of the first attempts in this direction was made by the Italian mathematician Luigi Fantappiè, around the middle of the last century. More or less simultaneously with Wheeler and Feynman, he developed similar considerations on the Dirac wave equation and applied himself to constructing a unitary theory of the physical and biological world on the basis of his results [14]. Fantappiè's unitary theory, centered on the concept of syntropy[16] as a generative principle of vital order in opposition to entropy, interpreter instead of the tendency to degradation and disorder of the inanimate world, has in fact provided many researchers, even in areas other than purely physical, with a philosophical basis for developing interpretative models of the macroscopic world in which causal (entropic) phenomena coexist with retrocausal (syntropic) phenomena giving rise to a *supercausal* vision of the natural world [2,3].

But it is mainly with the emergence of quantum biology [1,17] that, in recent decades, streams of research aimed at investigating the traces of retrocausal dynamics in highly complex areas, such as those dealing with neuroscience, have been developed. Given for acquired the existence of quantum structures in living systems,[17] the underlying idea is that all quantum objects (and therefore also those of biological interest) constantly find themselves in the need to make choices when faced with alternatives that are offered to them by *flows of information* that come from both the past and the future, according to the aforementioned transactional model; up to the hypothesis that biological retrocausality can make useful contributions to the investigation of the profound nature of conscious states. Many of the studies conducted in the last thirty years can be included in this specific area of research [15,17,23].

---

[16] See note 13.
[17] Studies on the magnetoreception of migratory birds, on photosynthesis and on physics of certain aspects of respiration can be cited as valid examples.

But there is a middle ground, still little explored.

If in the field of basic physics, quantum retrocausality (mainly in its broadest sense, which assumes that transactional dynamics take place in a dimension where the time variable does not exist) is an interpretative model that, although not universally accepted, has a solid theoretical basis, in areas characterized by higher levels of complexity, such as biological structures, the supposed existence of time-reversal dynamics is definitely much more speculative. What, therefore, would be useful to deepen is the possible presence of transactional mechanisms on the border between these two worlds: that is, where inanimate matter gives rise to life but, at the same time, complexity[18] is still *simple enough* to be investigated.

It is, therefore, in this spirit that we propose to interpret in quantum mechanical terms the results obtained through the IdEP-IdLA model, qualitatively described in paragraph 3. The aim is to verify whether it is possible, by this way, to reach the mutual accreditation of the transactional physical models, on the one hand, and of the telepoietic dynamics as the basis of abiogenesis, on the other.

5. <u>A possible transactional interpretation of telepoiesis</u>

With reference to the example proposed at the end of paragraph 2, let us look more closely at the formation of a target aggregate in accordance with the IdEP-IdLA model. Suppose that the chain begins with IdEP1 and that, in general, this particle can bind a second one among the four types present in the initial system: another IdEP1 or an IdEP2, an IdEP3, an IdEP4. Let us assume that, in view of the target aggregate, the IdEP1-IdEP2 pair is the correct one and that the other three combinations are wrong. If the binding energies between the pairs are very similar, none of the couplings is advantaged and the probability of establishing a bond is, for all pairs, equal or very close to 25%.[19]  If we repeat the procedure for the whole sequence of 100 IdEPs it is clear that an autopoietic process ($\eta = 0$) has a very low efficiency: it produces only one target aggregate every $1,6 \ 10^{60}$ chains!

But if the transactional narrative were valid, then things could go differently. In general, we can consider that bonds take place as a result of the energy emission by quantum emitters (electrons or atoms in an excited state) that are part of the IdEPs candidate to pair up each other. Then, we can regard IdEP1 as the absorber of electromagnetic perturbations generated by the motion of the binding particles that are on the IdEPs ready for coupling. In transactional terms, these would then generate offer waves, back and forth in time, to which the absorber would respond with as many response waves, also back and forth in time. The advanced effects of the absorber responses would produce, simultaneously with the emissions, as many advanced fields arriving at the IdEPs ready for coupling at the same time as the radiative fields are generated.

---

[18] Here, as elsewhere in the text, the term *complexity* is used in its traditional semantic sense to indicate highly structured aggregative forms.

[19] As has already been said, a simplified discussion is proposed here. The model is actually based on a substitution reaction in which the active IdEPs are provided by giver compounds: the energy that binds the IdEPs inside these reagents is assumed to be lower than the energy that will be released as a result of the subsequent aggregation processes (exothermic reaction).

Advanced and delayed fields, linked each other as a single entity, define the condition for incipient transactions, that is the possibility of energy transfer in accordance with Born probabilities. But which coupling will actually take place?

If IdEP1 and IdEP2 fields are such that the probability of IdEP1-IdEP2 coupling is a little more favoured than other possible couplings (that is, if the information of an aggregative code is conveyed by this way, exerting a not negligible statistical conditioning) then the formation of the right pair will be advantaged (i.e. more frequent). In this scenario, the interaction between IdEP1 and IdEP2 and the higher probability of IdEP1-IdEP2 pair formation is not the result of a deterministic mechanism but the outcome of transfer of information according to modalities that develop in an atemporal dimension.

Going forward, to the newly formed IdEP1-IdEP2 pair a third IdEP will have to be added, again selected among the four available types. Suppose the right particle is IdEP4. The same process repeats but, this time, the response waves depend on the IdEP1-IdEP2 pair. If IdEP1-IdEP2 and IdEP4 fields are likely to favour their bond (again, thanks to the conditioning of specific information living in a timeless dimension) in comparison to the formation of other possible triplets, then the group IdEP1-IdEP2-IdEP4 will have more opportunities to take place. And so on, with subsequent handshakes.

The whole process represents a mechanism pushing aggregate formation in the direction suggested by more or less strongly coded instructions, operating before the collapse of Shrödinger's wave function in what overmentioned researchers [19,20,21,22] consider the realm of possibilities.

In general, it can be assumed that, each time, a *significant segment* of the already formed IdEPs chain acts (in the domain of possibilities represented by the quantum substrate) on the next available IdEPs favouring, with greater or lesser intensity (depending on $\eta$ in IdEP-IdLA model), the aggregation of the right sequence (and therefore the construction of the target aggregate) according to the operation mode of a *source of information with memory*: the deeper the memory, the more efficient the code is. Under this respect, the coding factor $\eta$ of the IdEP-IdLA model would then combine two effects: the *average intensity of preference* (the addressing action exerted by the absorber) and the *depth of memory*, that is the number of IdEPs that serve to determine the right response wave.

The code may be more or less powerful but appreciable results, from an abiogenetic perspective, can also be obtained with relatively low values of $\eta$, as seen in paragraph 2. On the other hand, to the extent that the proposed theory can contribute to the understanding of the first onset of life, it seems reasonable to assume that the values of $\eta$ in a telepoietic context should be quite far from unity. Only in this way, in fact, can be given reason for the fact that the spontaneous formation of structures of biological interest is, however, quite rare and rather difficult to be reproduced. Certainly, if telepoiesis exists in nature, it can only coexist with causal dynamics, in a classically deterministic sense, that represent disturbing elements, like a sort of background noise.

Crediting as realistic, at the microscopic level, the dynamics just hypothesized would, therefore, confirm the telepoietic interpretation of the results proposed by the IdEP-IdLA model as sustainable: in the domain of possibilities (which live outside spacetime) are most likely implemented some options that in spacetime give rise to irreversible effects that we read as finalized. At the same time, to the extent that the results of IdEP-IdLA model don't leave room for different assumptions to justify the uncompensated neghentropy $\sigma$ other than the existence of a not conventional flow of information, the model itself confirms the existence of quantum phenomena of the transactional type. Far from being a vicious circle, this mutual aid seems to give a certain solidity to the whole scenario. In essence, what is prefigured is that, in a flow of macroscopic events that develop irreversibly in cosmological time, at the microscopic level, before the collapse of wave functions describing quantum phenomena, nature has a means of *conveying soft targeted information*.

All this is not a demonstration but, at least, a strong indication, a viable working hypothesis perhaps to be deepened in those open systems, dominated by constructive chaos, which are the dissipative structures already mentioned and which Prigogine indicates as the privileged place of abiogenesis [24]. The model itself, as already mentioned at the end of paragraph 3, seems to suggest this [7]. Clearly, if telepoiesis were really part of the grammar necessary for the complete representation of the natural world then, probably, it would not be reasonable to consider its effects confined to abiogenesis alone. Perhaps, telepoiesis has continued (and still does) to exert its building action in the natural world, even after the first beginning of life. Perhaps, it permeates the entire biosphere, thus justifying on the one hand the evident tension towards a growing complexity of living organisms and, on the other, the tortuous path that evolution follows as a result of the powerful disruptive actions induced by entropic phenomena. In this perspective, the exploitation of the aforementioned supercausal interpretations of the natural world is certainly a possibility. The whole phenomenological world could thus be seen as a kind of fabric in which a robust warp of events, governed by causes firmly anchored in the past, intertwists with a tenuous weft of teleonomic events [8].

6. <u>Conclusions</u>

The Idep-IdLA model, applied to abiogenesis, prefigures the existence, in nature, of tele-causal dynamics whose foundation can be sought in the domain of possibilities subject to actualization, living in the realm of quantum substratum.

Conversely, insofar as transactional interpretation of quantum mechanics is not universally accepted by physicists, the clues provided by the model bring elements in favour of the real possibility of its ontological existence at the level of basic physics. It is evident that the reasoning developed here leaves many questions unanswered. Certainly, if telepoiesis exists, then it lives in a dialectical relationship with the classical causal dynamics while conditioning the natural world in an important and still substantially unknown way. But it would be quite futile to search for its way of acting by means of a traditional approach since, by its nature, any possible telepoietic dynamics would escape experimental investigation as we usually practice it.

Rather, it would be necessary to look for its effects *a posteriori* in the macroscopic world, where open systems far from equilibrium tend, sometimes, to form highly organized structures, maximizing entropy production. In this perspective, the entire evolutionary history could be critically analysed through the telepoietic key of reading.

<u>Acknowledgments</u>

The authors are grateful to Enrico Giannetto (Università degli Studi di Bergamo) for encouraging this research and to Ruth E. Kastner (University of Maryland) for her valuable and irreplaceable suggestions that have allowed to correctly frame the work within the transactional interpretation of quantum mechanics.

<u>REFERENCES</u>

[1]    Al-Khalili J., McFadden J. (2014), *Life on the edge. The coming of age of quantum biology*, Published in Italy by Bollati Boringhieri in 2015 - ISBN: 978-88-339-2300-0

[2]    Arcidiacono G., Arcidiacono S. (1991), *Sintropia, entropia, informazione*, Di Renzo Editore, Roma – ISBN: 88-83231-51-1

[3]    Arcidiacono S. (1993), *L'evoluzione dopo Darwin*, Di Renzo Editore, Roma – ISBN: 88-83231-11-2

[4]    Cocchi V, Morandi R. (2021), *Mathematical modeling for the simulation of aggregative processes*, Entropie (ISTE Editions) DOI 10.21494/ISTE.OP.2021.0667.

[5]    Cocchi V, Morandi R. (2022), *Mathematical simulation of aggregative processes: generalization of the IdEP-IdLA model* – Entropie (ISTE Editions), DOI: 10.21494/ISTE.OP.2022.0858.

[6]    Cocchi V, Morandi R. (2021), *Use of the IdEP-IdLA model for the study of aggregative processes in closed systems*, Entropie (ISTE Editions) – DOI: 10.21494/ISTE.OP.2021.0741.

[7]    Cocchi V, Morandi R. (2021), *Aggregative processes in open systems: simulations and detailed thermodynamic analysis by means of the IdEP-IdLA model,* Entropie (ISTE Editions) – DOI: 10.21494/ISTE.OP.2021.0742.

[8]    Cocchi V., Morandi R. (2021), *The IdEP-IdLA model and the biochemistry of aggregative processes: the Arianna's conjecture,* Entropie (ISTE Editions) – DOI:   10.21494/ISTE.OP.2021.0755.

[9]    Cramer J.C. (1980), *Generalized Absorber Theory and the Einstein-Podolsky-Rosen Paradox*, Physical Review D, 22(2): 362–376. DOI:10.1103 Phys Rev D.22.362

[10]   Cramer J.C. (1986), *The Transactional Interpretation of Quantum Mechanics*, Reviews of Modern Physics, 58(3): 647–687. DOI:10.1103 RevModPhys.58.647

[11]   Cramer J.C. (1988), *An Overview of the Transactional Interpretation of Quantum Mechanics*, International Journal of Theoretical Physics, 27(2): 227–236.  DOI: 10.1007 BF00670751

[12]   Cramer J.C. (2016), *The Quantum Handshake*, Cham: Springer International Publishing. DOI:10.1007/978-3-319-24642-0

[13]   Evans P.W. (2022), *Retrocausality in Quantum Mechanics*, *The Stanford Encyclopedia of Philosophy* (Winter 2022 Edition), Edward N. Zalta & Uri Nodelman (eds.), URL = <https://plato.stanford.edu/archives/win2022/entries/qm-retrocausality/>.

[14]   Fantappiè L. (1944), *Principi di una teoria unitaria del mondo fisico e biologico,* Reprinted in 1993 by Di Renzo Editore, Roma – ISBN: 88-86044-11-9

[15]   Hameroff S., Penrose R. (2014), *Consciousness in the universe: a review of the 'Orch OR' theory*, in Physics of Life Reviews, vol. 11, n. 1, 2014, pp. 39-78,  Bibcode: 2014 PhLRv..11...39H, DOI:10.1016/j.plrev.2013.08.002, PMID 24070914

[16]   Hill T.L. (1986), *An Introduction to Statistical Thermodynamics*, *Paragraph 8-4*, Dover Pubblication Inc, New York – ISBN: 0-486-46742-2

[17]   Hoffmann P.M. (2012), *Life's racket. How molecular Machines esxtract order from chaos*, Published in Italy by Bollati Boringhieri in 2014 – ISBN: 978-88-339-2505-9

[18]   Kastner R.E. (2012), *The Transactional Interpretation of Quantum Mechanics: the Reality of Possibility*, Cambridge University Press (1st edition)

[19]   Kastner R.E. (2016), *Is There Really "Retrocausation" in Time-Symmetric Approaches to Quantum Mechanics?* – arXiv:1607.04196

[20]   Kastner R.E., Kauffman S., Epperson M. (2018), *Taking Heisenberg's Potentia Seriously*, International Journal of Quantum Foundation – Volume 4, Issue 2, pages 158-172

[21]  Kastner R.E. (2021), *The Relativistic Transactional Interpretation and Spacetime Emergence*, Preprint of [22]

[22]  Kastner R.E. (2022), *The Transactional Interpretation of Quantum Mechanics: a Relativistic Treatment,* Cambridge University Press (2nd edition), Chapter 8

[23]  King C.C. (2003), *Chaos, Quantum-transactions and Consciousness, Neuro Quantology*, Vol. 1(1):129-162; Syntropy 2007, 1, pag.29-48 ISSN: 1825-7968 www. sintropia.it

[24]  Kondepudi D., Prigogine I. (1998), *Modern Thermodynamic, from Heat Engines to Dissipative Structures*, Jhon Wiley & Sons – ISB: 0-471-97393-9.

[25]  Tresoldi F. (2008), *Irreversibilità quantistica. Un'analisi del ruolo del tempo in meccanica quantistica,* Tesi di dottorato Università di Bergamo, https://aisberg.unibg.it/handle/10446/41

[26]  Wheeler J.A., Feynman R.P. (1945), *Interaction with the Absorber as the Mechanism of Radiation*, *Reviews of Modern Physics*, 17(2–3): 157–181. DOI:10.1103 Rev. Mod Phys.17.157

[27]  Wheeler J.A. e Feynman R.P. (1949), *Classical Electrodynamics in Terms of Direct Interparticle Action*, *Reviews of Modern Physics*, 21(3): 425–433. DOI: 10.1103 Rev: Mod Phys.21.425